\documentclass[final,5p,times,twocolumn]{elsarticle}

\usepackage{graphicx}

\usepackage{amsmath}
\usepackage{amssymb}
\usepackage{bm}
\usepackage{float}
\usepackage{color}
\usepackage{listings}    

\biboptions{comma,square}

\journal{Computer Physics Communications}

\restylefloat{figure}

\definecolor{dkgreen}{rgb}{0,0.6,0}
\definecolor{MyDarkBlue}{rgb}{0.0,0,0.7}

\begin{document}

\begin{frontmatter}



\title{High precision series solution of differential equations:\\
Ordinary and regular singular point of second order ODEs.}


\author[ify]{Amna Noreen}
\ead{Amna.Noreen@ntnu.no}

\author[ify]{K{\aa}re Olaussen}
\ead{Kare.Olaussen@ntnu.no}

\address[ify]{Institutt for fysikk, NTNU}

\begin{abstract}
A subroutine for very-high-precision numerical solution
of a class of ordinary differential equations is provided.
For given evaluation point and equation parameters 
the memory requirement scales linearly with  precision $P$,
and the number of algebraic operations scales roughly linearly
with $P$ when $P$ becomes sufficiently large.
We discuss results from extensive tests of the code, and
how one for a given evaluation point and equation parameters may
estimate precision loss and computing time in advance.
\end{abstract}


\end{frontmatter}





{\bf PROGRAM SUMMARY}

\begin{small}
\noindent
{\em Manuscript Title: } High precision series solution of differential equations:
Ordinary and regular singular point of second order ODEs.\\
{\em Authors:} Amna Noreen, K{\aa}re Olaussen\\
{\em Program Title:} seriesSolveOde1\\
{\em Journal Reference:}                                      \\
{\em Catalogue identifier:}                                   \\
{\em Licensing provisions:} none                              \\
{\em Programming language:} C++                               \\
{\em Computer:} PC's or higher performance computers          \\
{\em Operating system:} Linux and MacOS                       \\
{\em RAM:} Few to many megabytes (problem dependent)         \\
{\em Number of processors used:} 1                            
{\em Classification:} 2.7 Wave functions and integrals,
4.3 Differential equations.\\
{\em External routines/libraries:}
CLN -- Class Library for Numbers~\cite{CLN}
built with the GNU MP library~\cite{GMP},
and GSL -- GNU Scientific Library~\cite{GSL} 
(only for time measurements).\\
{\em Subprograms used:}                                   
The code of the main algorithm is in the file {\tt seriesSolveOde1.cc},
which \lstinline!#include! the file {\tt checkForBreakOde1.cc}. 
These routines, and programs using them, must \lstinline!#include!
the file {\tt seriesSolveOde1.cc}.\\ 
{\em Nature of problem:}
The differential equation
\begin{equation}
     -s^2\left(\frac{d^2}{dz^2}  + 
     \frac{1-\nu_+ - \nu_-}{z}\frac{d}{dz} + 
     \frac{\nu_+ \nu_-}{z^2} \right)\psi(z) + 
   \frac{1}{z}\sum_{n=0}^{N} \text{v}_n z^n\,\psi(z) = 0,
   \label{ODE}
\end{equation}
is solved numerically to very high precision. The evaluation point $z$ and some or all
of the equation parameters may be complex numbers; some or all of them may be represented
exactly in terms of rational numbers.\\
{\em Solution method:}\\
The solution $\psi(z)$, and optionally $\psi'(z)$, is evaluated at the point $z$ by executing the recursion
\begin{align}
  A_{m+1}(z) &= \frac{s^{-2}}{(m+1+\nu -\nu_+)(m+1+\nu-\nu_-)} \sum_{n=0}^N V_n(z)\,A_{m-n}(z),
  \label{RecursionRelation}\\
  \psi^{(m+1)}(z) &= \psi^{(m)}(z) + A_{m+1}(z),\label{SummingUpPsi}
\end{align}
to sufficiently large $m$. Here $\nu$ is either $\nu_+$ or $\nu_-$, and $V_n(z) = \text{v}_n\,z^{n+1}$.
The recursion is initialized by 
\begin{align}
  A_{-n}(z) &= \delta_{n0}\,z^\nu,\quad\text{for $n=0, 1,\ldots,N$}\\
  \psi^{(0)}(z)&= A_0(z).
\end{align}
{\em Restrictions:}
No solution is computed if $z=0$, or $s=0$, or  if $\nu=\nu_-$ 
(assuming $\text{Re}\, \nu_+ \ge \text{Re}\, \nu_-$) with $\nu_+ - \nu_-$
an integer, except when $\nu_+ - \nu_- = 1$ and $\text{v}_0 = 0$ 
(i.e.~when $z$ is an ordinary point for $z^{-\nu_-}\,\psi(z)$).
\\
{\em Running time:}
  On an few years old Linux PC, evaluating the ground state wavefunction of
  the anharmonic oscillator (with the eigenvalue known in advance),
  cf. equation (\ref{AnharmonicOscillator}), at $y=\sqrt{10}$ to $P=200$ decimal digits
  accuracy takes about $2\;\text{milliseconds}$, to $P=100\,000$ decimal digits accuracy takes
  about $40\;\text{minutes}$.

\end{small}

\section{Introduction}

Modelling and analysis of many problems in science and engineering involves
the solution of ordinary differential equations, sometimes in the domain of complex numbers.
For practical use such solutions must usually be computed numerically at some stage.
In some cases it may be necessary, useful, or interesting
to do this to much higher precision than provided by standard equation
solvers (or routines for evaluating standard functions).

Some examples of cases where numerical calculations have been used
to inspire or check analytic conjectures and proofs are the works
by Bender and Wu \cite{BenderWu} and Zinn-Justin and
Jentschura \cite{ZinnJustin_Jentschura}. With access to very accurate
numerical results the opportunities for such explorations increases.
Access to essentially exact results are also useful for analyzing the behaviour
of approximation schemes, as in the work by Bender {\em et al} \cite{Bender_etal}
and more recently by Mushtaq {\em et al} \cite{AAK}.

We have implemented and investigated the algorithm~(\ref{RecursionRelation}, \ref{SummingUpPsi})
for solving equation~(\ref{ODE}) to very high precision, and believe the
C++ function {\tt seriesSolveOde1} may be of use or interest to others.
An early version of this code has been used to solve eigenvalue
problems like the anharmonic oscillator and the double well potential,
\begin{align}
   &\left[-\frac{d^2}{dy^2} + y^4\right]\psi(y) = \varepsilon_n\,\psi(y), \label{AnharmonicOscillator}\\
   &\left[-s^2\frac{d^2}{dy^2} + \left(1-y^2\right)^2\right]\psi(y) = \varepsilon_{n\sigma}\,\psi(y) \label{DoubleWell},
\end{align} 
to very high precision. In reference \cite{AAKI} the ground state eigen\-energy $\varepsilon_0$
of (\ref{AnharmonicOscillator}) was found to $1^+$ million decimal digits precision,
the excited state $\varepsilon_{50\,000}$ was solved to $50\,000^+$ decimals,
and the lowest even, $\varepsilon_{0+}$, and
odd parity, $\varepsilon_{0-}$, eigenvalues of (\ref{DoubleWell}), with $s=1/50\,000$,
was found to $30\,000^+$ decimals. Equations (\ref{AnharmonicOscillator}, \ref{DoubleWell})
are transformed to the form (\ref{ODE}) by introducing $z=y^2$, leading to
$\nu_-=0$ and $\nu_+ = \frac{1}{2}$ and $\text{v}_2 = \frac{1}{4}$. 
This further gives $\text{v}_0= -\frac{1}{4} \varepsilon_n$ for equation (\ref{AnharmonicOscillator}),
and $\text{v}_0 = \frac{1}{4}(1 - \varepsilon_{n\sigma})$, $\text{v}_1 = -\frac{1}{2}$ for
equation (\ref{DoubleWell}). I.e., the eigenvalue parameter enters in the coefficient $A_0(z)$
of equation (\ref{SeriesExpansion}).

The eigenvalue condition for these problems is that
the wave function should vanish as $y\to \pm\infty$, a condition which cannot be
imposed numerically. However, an asymptotic analysis of the behaviour of the wavefunction
as $y\to \pm\infty$ allows us to replace it with an 
equivalent Robin boundary condition at finite $y$. The latter cannot be computed exactly,
but to sufficient accuracy for any desired precision. In fact, if we make $y$ large enough
it suffices to use a Diriclet boundary condition.

In reference \cite{ANKO} it was demonstrated
that the wavefunction normalization integrals can be computed to comparable precision,
again using an early version of our code.

In the rest of this paper we provide examples of how this code can be used,
and some analysis of its behaviour. We do not focus on specific areas
of applications, but would like to mention that very-high-precision
computations of Green functions and functional determinants are possible applications.
The code can evaluate $\psi(z)$ for complex values of $z$, and
allow for complex parameters in the differential equation.

\section{Basic use}

The function \lstinline!seriesSolveOde1! is declared as
\lstset{
  language=C++,
  title={\normalsize\em Function declaration},
  morekeywords={OdeParams, OdeResults, cl_N, cl_I, float_format_t}
}
\begin{lstlisting}
bool seriesSolveOde1(OdeResults& results, const OdeParams& params)
\end{lstlisting}
The function returns \lstinline!true! if the calculation completed normally
and \lstinline!false! if the calculation was aborted. Function arguments and
options are collected in a structure \lstinline!OdeParams!, with the
function value $\psi(z)$, optionally $\psi'(z)$, and various diagnostic results
returned in a structure \lstinline!OdeResults!. The definitions of these
structures are listed in the Appendix at the end of this paper.
A code snippet illustrating the use of \lstinline!seriesSolveOde1! is
\lstset{
  language=C++,
  title={\normalsize\em Example use},
  morekeywords={OdeParams, OdeResults, cl_N, cl_I, float_format_t}
}
\begin{lstlisting}
  // Better start from default parameters
  OdeParams params = defaultOdeParams;
  // Change parameters as needed
  params.prec = float_format(500);
  params.z = complex(27/2, 43/7);
  params.dAlso = true;
  OdeResults results;
  // NB! Must allocate space for $\psi(z)$ a $\psi'(z)$
  cl_N fu[2]; 
  results.fu = fu;
  if(seriesSolveOde1(results, params)){
    cout << results.fu[0] << endl;
    cout << results.fu[1] << endl;
  }
\end{lstlisting}

\section{Computational accuracy}

Our solution is found by a {\em brute force\/} summation,
\begin{equation}
    \psi(z) = \sum_{m=0}^{\infty}\,a_m\,z^{\nu+m} \approx \sum_{m=0}^{\cal M}\,a_m\,z^{\nu+m} \equiv \sum_{m=0}^{\cal M}\,A_m(z).
    \label{SeriesExpansion}
\end{equation}
Since equation (\ref{ODE}) has no singularities in
the finite $z$-plane, except perhaps $z=0$,
the sum in guaranteed to have an infinite radius of convergence.
However, intermediate terms in the sum may be very large although
the final result is small; hence there may be huge cancellations,
leading to significant loss of accuracy when $z$ is large.

The computations are performed with high-precision floating point
numbers, with precision regulated by the parameter
\lstinline!params.prec! as shown in the code snippet above. The
value given is the intended precision $P$ in decimal digits, but since
memory for floating point numbers is allocated internally
(in CLN at high precision) in chunks of 64~bits the actual precision
is usually somewhat higher --- increasing in steps
of about $19 \approx 64\cdot \log 2 /\log 10$ decimal digits.

Denote the actual precision used in computation by $M$. I.e.,
each non-zero floating point number $x$ is represented as $(-1)^s 2^m\,f$,
with the mantissa $f$ ($\frac{1}{2} \le f < 1$) given to a precision of
$M$ bits. This means that the potential roundoff error in $x$ is $2^{m-M-1}$.
I.e, if the largest term $A_m(z)$ in the sum (\ref{SeriesExpansion}) has the
representation $(-1)^s 2^{\bar{A}}\,f$ it may contribute a roundoff error
$2^{\bar{A}-M-1}$ to $\psi$. Due to the recursion relation (\ref{RecursionRelation})
roundoff errors may be further amplified (or partially cancelled),
but it is a reasonable hypothesis to use $\bar{A}$ for an estimate of
the evaluation error.

\begin{figure}[H]
\begin{center}
\includegraphics[clip, trim = 6ex 6ex 6ex 4ex, width=0.483\textwidth]{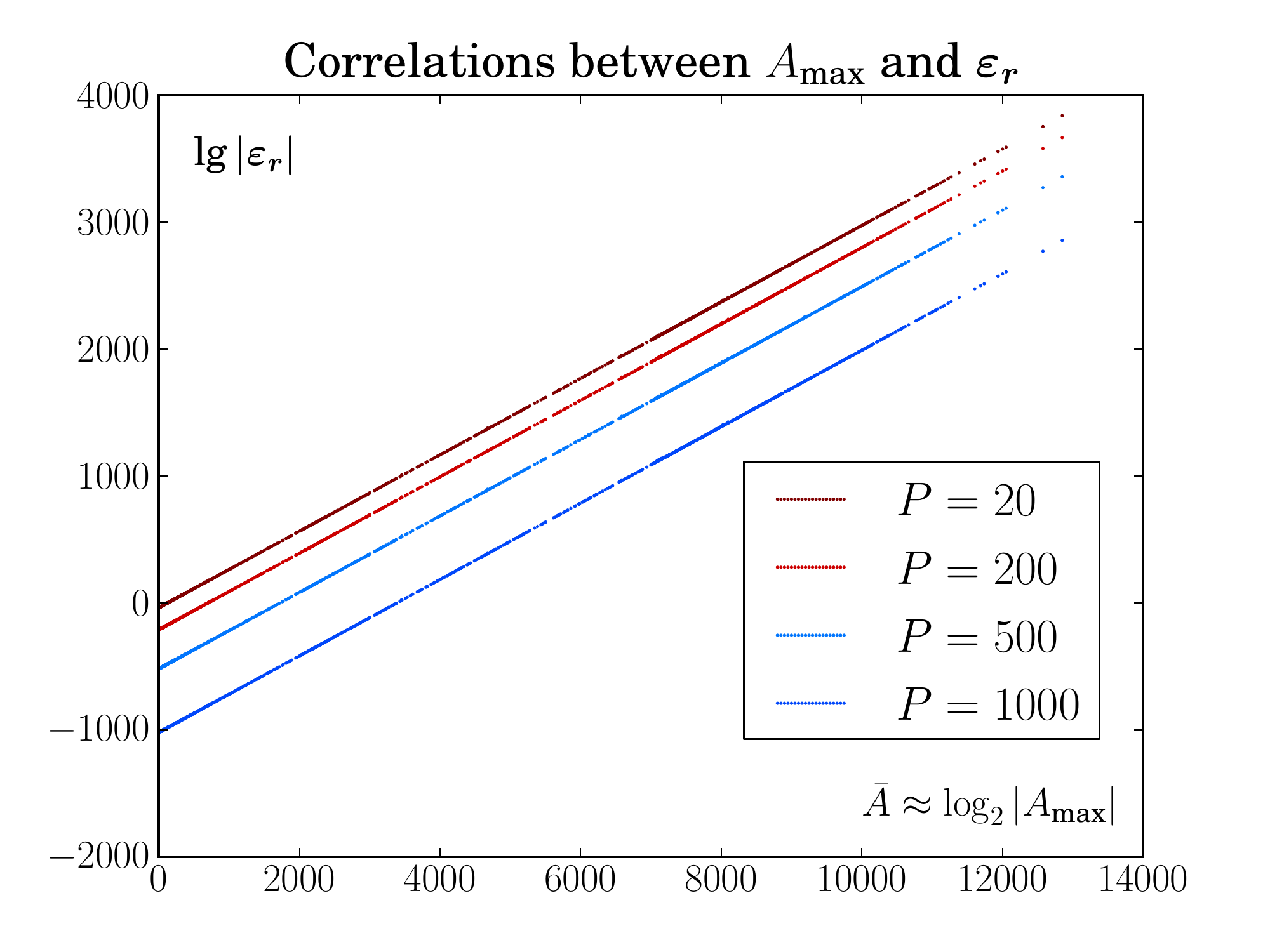}
\end{center}
\caption{The real evaluation error, $\varepsilon_r \equiv \vert \psi_P(z) -\psi_E(z)\vert$, caused by roundoff
is strongly correlated with the largest term $A_{\text{max}}$ in the series expansion. Here $\psi_P(z)$ is
the value obtained when evaluating the series to the intended precision of $P$ decimal digits, while
$\psi_E(z)$ is the ``exact'' value (here actually the value obtained for $P=1040$).}
\label{maxA_realError_correlation}
\end{figure}

We have tested this hypothesis by running a
large number ($200\,000$) of evaluations with randomly chosen para\-meters,
and investigated the correlation between $\bar{A}$ and the real evaluation
error $\varepsilon_r \equiv \vert \psi_P(z) -\psi_E(z)\vert$ for various
precisions $P$. Here $\psi_E(z)$, representing the exact value, is found by
doing the computation with a precison $P_E$  reasonably larger than
all the others. As can been seen in figure~\ref{maxA_realError_correlation}
the correlation between $\bar{A}$ and $\varepsilon_r$  is good compared to
the accuracies in question.

For diagnostic purposes the value of $\bar{A}$ is returned
by \lstinline!seriesSolveOde1! in the variable
\lstinline!results.maxAExponent!. The corresponding value
for $\psi'(z)$ is returned in the variable
\lstinline!results.maxAdExponent! (the values of $m$ where
the maxima occur are also returned). Based on these values
and the actual precision $M$ the estimated errors in decimal digits
are returned in the variables \lstinline!results.lgErrorF!
and \lstinline!results.lgErrorFd!. Their exact values are
$(\bar{A} - M)\log 2/\log 10 + G$ for $\psi$, and
$(\bar{A}' - M)\log 2/\log 10 + G'$ for $\psi'$,
where the numbers $G=4.30$ and $G' = 3.02$ are empirically
choosen ``guard digits'' to avoid underestimating the error (too often).

\begin{figure}[H]
\begin{center}
\includegraphics[clip, trim = 8ex 6ex 9ex 5ex, width=0.483\textwidth]{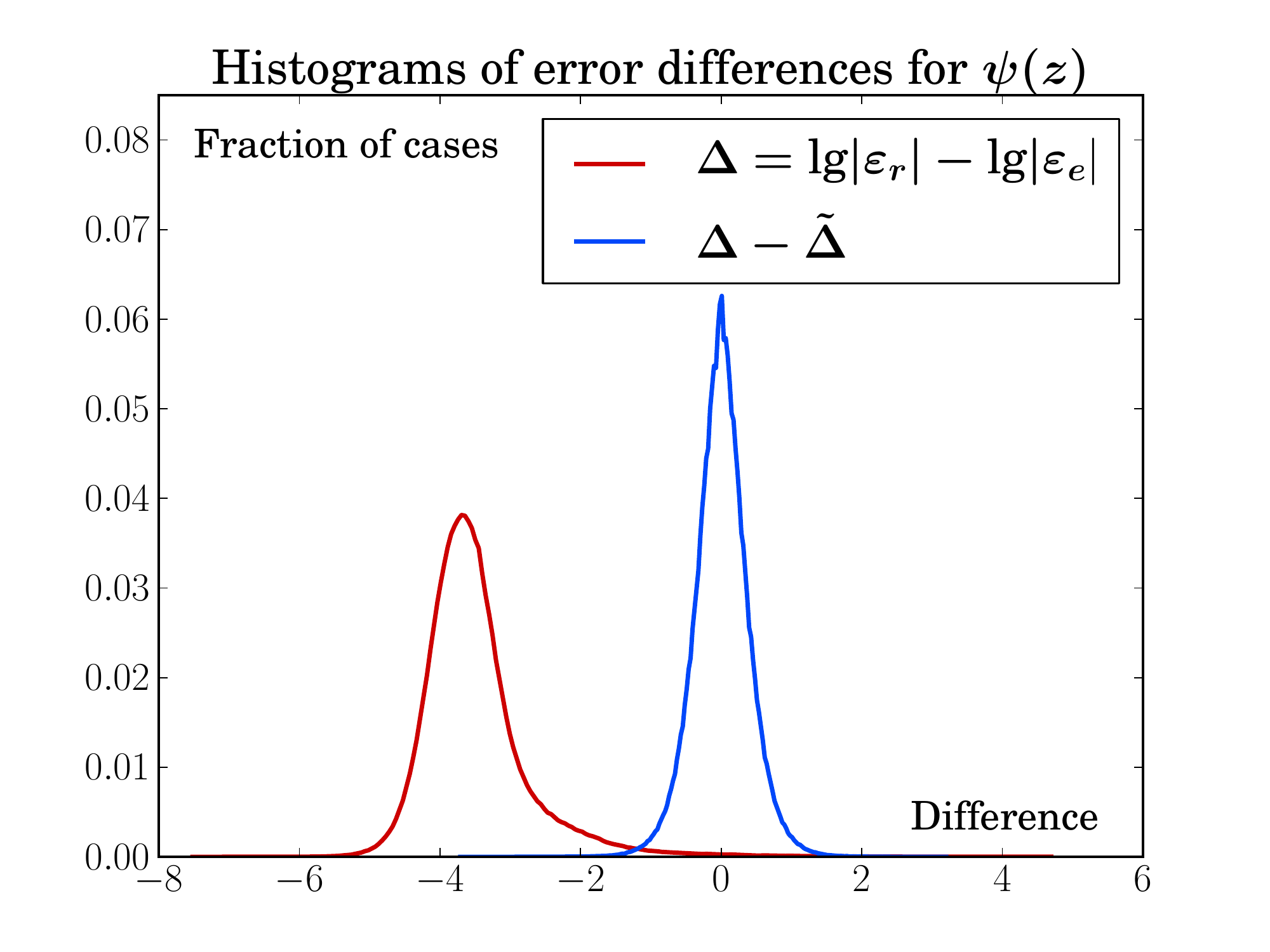}
\end{center}
\caption{The curve for $\Delta= \lg\vert\varepsilon_r\vert - \lg\vert\varepsilon_e\vert$ shows a histogram
of differences between the real evaluation error $\varepsilon_r$, and an estimate based on the largest
term $A_{\text{max}}$ in the sum (\ref{SeriesExpansion}). The histogram does not depend on computational precision $P$, but
individual differences (with fixed $z$ and equation parameters) varies with $P$ in a fluctuating manner. The curve for
$\Delta - \tilde{\Delta}$ shows how differences computed at precision
$P=20$ ($\tilde{\Delta}$) correlates with those computed at $P = 200, 500, 1000$.}
\label{RealAndEstimatedErrors}
\end{figure}

These estimates are accurate to a handful of decimal digits as shown by
the $\Delta= \lg\vert\varepsilon_r\vert - \lg\vert\varepsilon_e\vert$ histogram
in figure~\ref{RealAndEstimatedErrors}. We have found such histograms
to be independent of computational precision $P$, and (with the choosen value of
$G-G'$) also the same for $\psi(z)$ and $\psi'(z)$. The histogram is taken over
200\,000 evaluations with $N$ choosen randomly between $1$ and $4$, the real and imaginary
parts of $s$ randomly from the set $\{ -1, -\frac{1}{3}, \frac{1}{3}, 1 \}$,
the real and imaginary parts of $\nu_{\pm}$ randomly between $-10$ and $10$, the real and
imaginary parts of $\text{v}_n$ randomly between $-5$ and $5$, and the real and
imaginary parts of $z$ randomly between $-20$ and $20$.

The difference $\Delta$ is caused by an essentially unpredictable roundoff error,
amplified by a recursion relation which depends on $z$ and parameters of the
differential equation. As can be seen from figure~\ref{RealAndEstimatedErrors} the
ratio between the real and estimated errors varies between almost
$10^5$ and $10^{-8}$. Although this variation is large it is still
less than the discrete steps by which the actual precision is increased.

We have investigated how the difference $\Delta$ computed at different precisions
(but for the same evaluation point $z$ and equation parameters) are correlated. This is shown in
the $\Delta -\tilde{\Delta}$ histogram in figure~\ref{RealAndEstimatedErrors}, where
$\tilde{\Delta}$ refer to a computation with intended precision $P=20$, and $\Delta$
to computations with $P = 200$, $500$, and $1\,000$ (where each $P$ gives the same
looking histogram). As can be seen the correlations are stronger than
for a single $\Delta$, but there are still wide tails.

In conclusion, the largest term $A_{\text{max}}$ in the series~(\ref{SeriesExpansion}),
or the associated integer $\bar{A}$, provides a good estimate of the evaluation
error, but the real error may still differ by several orders of magnitude. As somewhat
better empirical estimate can be obtained by first computing the
real error at low $P$ (where it is computationally inexpensive) and assuming
\begin{equation}
    \lg \varepsilon_E = \lg \tilde{\varepsilon}_r - (M - \tilde{M})\log 2/\log 10 + 2.
\end{equation}
Here $\varepsilon_r$ is the real error at $\tilde{M}$ bits of actual precision, with $\varepsilon_E$ 
the estimated error at $M$ bits of actual precision.

\section{Comparison with exactly known Wronski determinant}

In the previous section we assumed that \lstinline!seriesSolveOde1! would compute
accurate results at large intended precision $P$, but this was not really
verified. One check is to compare its results with the exactly known Wronski determinant,
\begin{equation}
    W(z) = \psi_{\nu_+}(z)\,\psi'_{\nu_-}(z) - \psi_{\nu_-}(z)\,\psi'_{\nu_+}(z) 
    = (\nu_- - \nu_+) z^{\nu_+ + \nu_- -1}.
\end{equation}

\begin{figure}[H]
\begin{center}
\includegraphics[clip, trim= 6.5ex 6ex 9ex 5ex, width=0.483\textwidth]{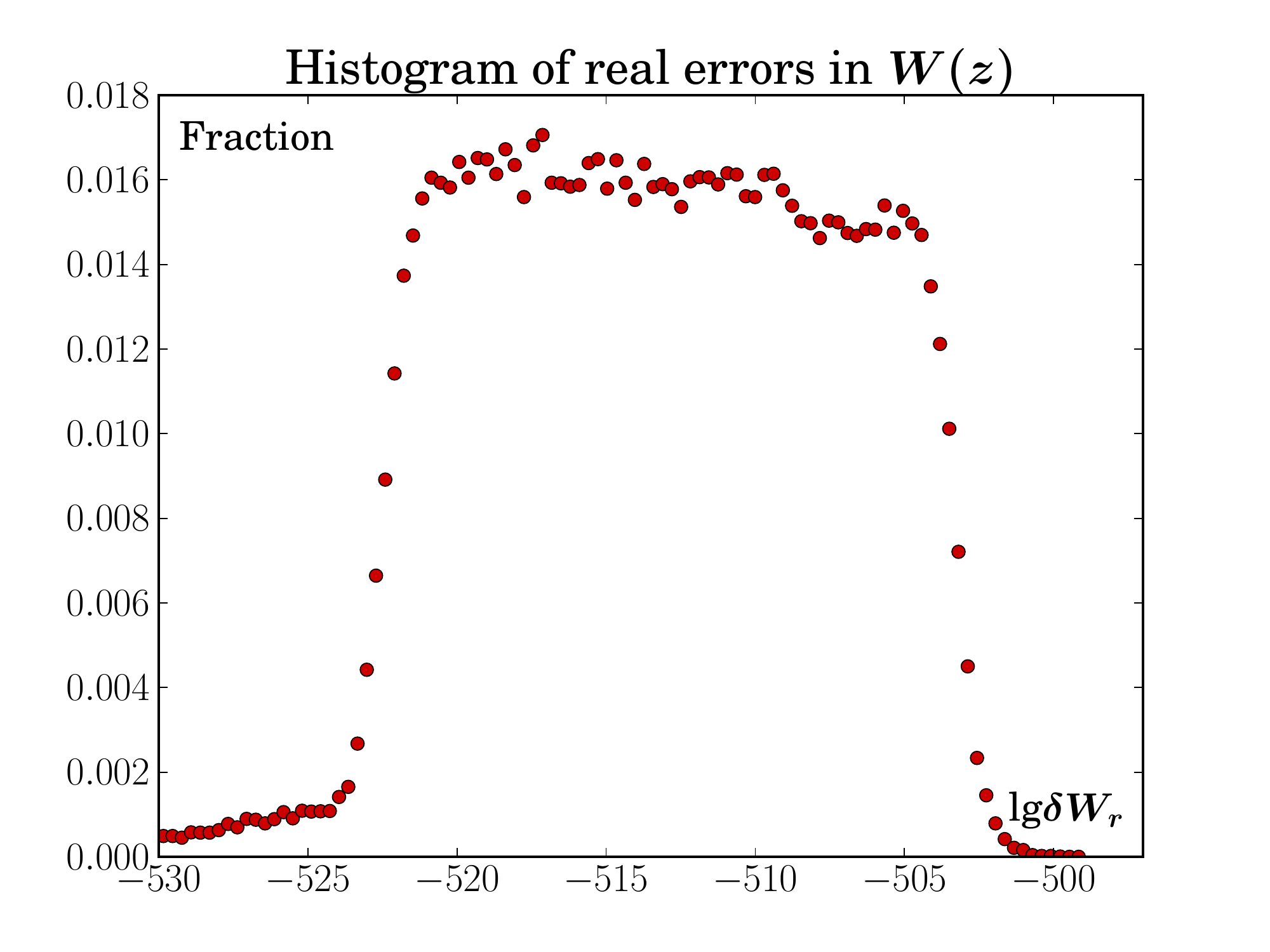}
\end{center}
\caption{The real errors
$\delta W_r \equiv \vert W^{\text{exact}}(z) - W^{\text{num}}(z) \vert$
when aiming to compute the Wronski determinant $W(z)$ to an absolute
accuracy of $10^{-500}$. The histogram is taken over $150\,000^+$
random evaluation points $z$ and equation parameters.
}
\label{wronskiRealErrors}
\end{figure}

\begin{figure}[H]
\begin{center}
\includegraphics[clip, trim = 8ex 6ex 9ex 5ex, width=0.483\textwidth]{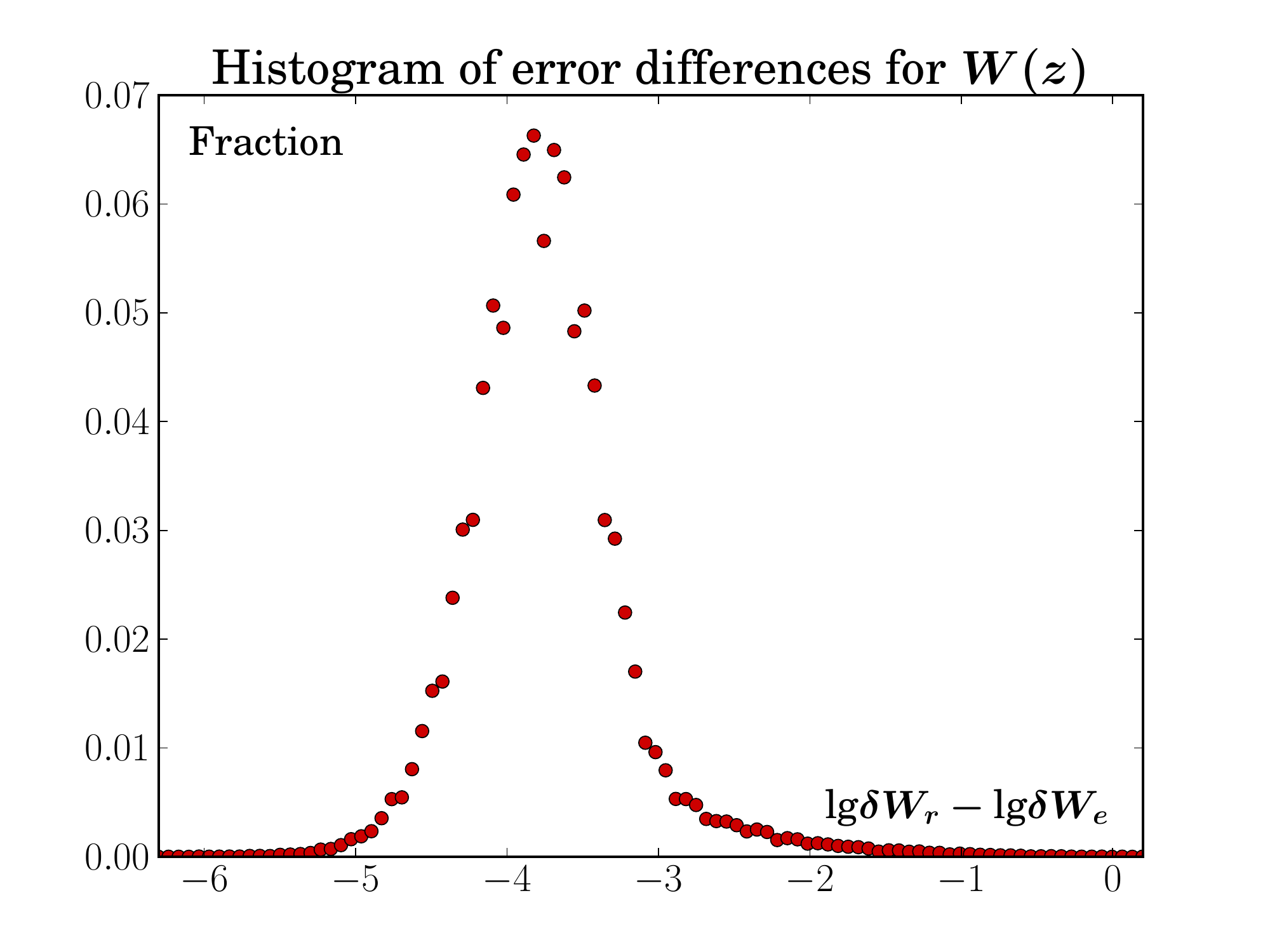}
\end{center}
\caption{Histogram of differences between the real error $\delta W_r$ in the numerically
computed Wronski determinant, and the estimated error $\delta W_e$ based
on (\ref{estimatedWronskiError}).
}
\label{wronskiErrorDifferences}
\end{figure}

The numerically computed determinant is estimated to have an error of magnitude
\begin{equation}
    \delta W_e = \max \left\{ 
      \vert\psi_{\nu_+}\vert\, \delta\psi'_{\nu_-},
      \vert \psi_{\nu_-}\vert\, \delta\psi'_{\nu_+}, 
      \vert \psi'_{\nu_-}\vert\, \delta\psi_{\nu_+}, 
      \vert \psi'_{\nu_+}\vert\, \delta\psi_{\nu_-}
    \right\},
    \label{estimatedWronskiError}
\end{equation}
where f.i. $\delta\psi'_{\nu_-}$ is the estimated magnitude of error in $\psi'_{\nu_-}$
(all quantities evaluated at $z$). An estimate of the loss of precision can be made
by a calculation at low(er) $P$, and used to choose the appropriate value of
\lstinline!params.prec! for a desired final precision in $W(z)$.
Figure~\ref{wronskiRealErrors} shows a histogram of how this works, tested on a
large number of random evaluation points and equation parameters. In most
cases the real precision is reasonbly close (always better) than the
one aimed for, but sometimes it turns out to be much better. This may occur
when the low precision calculation overestimates
the magnitude of $\psi(z)$ or $\psi'(z)$. However, as shown in
figure~\ref{wronskiErrorDifferences} the real error
$\delta W_r$ in the numerically computed determinant is always reasonably close to
the final estimate $\delta W_e$ based on equation (\ref{estimatedWronskiError}) with
all quantities computed at high precision.

The conclusion is that the calculated Wronski determinants are correct
within the expected accuracies, at least for computations at
sufficiently high precision (how high may depend on the evaluation point $z$
and equation parameters).

\section{{\em A priori\/} accuracy estimates}

Although \lstinline!seriesSolveOde1! monitors the largest term in
the sum~(\ref{SeriesExpansion}) to estimate the accuracy
of the computed results, it is desirable to predict the behaviour
of the sum in advance.
One way to do so is by analysing
the behaviour of the recursion relation~(\ref{RecursionRelation}).
Quite detailed and interesting results can be found in simple cases, but the analysis becomes
unmanageable in general. We instead make the hypothesis that the terms in
the sum~(\ref{SeriesExpansion}) for large $z=x\,\text{e}^{\text{i}\varphi}$ is strongly peaked
(in absolute value) around some $m=\bar{m}$, and that there exist
values of $\varphi$ for which there are little
cancellation between the large terms. I.e., we assume that
\begin{equation}
  \mathop{\text{max}}_{\varphi} \left|\, \psi(x\,\text{e}^{\text{i}\varphi}) \,\right| \approx 
  \vert a_{\bar{m}} \vert\,x^{\nu+\bar{m}},
  \label{BasicHypothesis}
\end{equation}
for positive $x$. We may use the WKB-approximation to estimate the left hand side.
For analytic treatment we first neglect the slowly varying algebraic prefactor
of the WKB-approximation, and a similar correction to the relation~(\ref{BasicHypothesis}).
Such corrections can be included in a numerical implementations.

Define, for positive $u$,
\begin{equation}
     S(u) = \mathop{\text{max}}_{\varphi} \log{\left|\psi(\text{e}^{u+\text{i}\varphi})\right|}\,
     \approx \mathop{\text{max}}_{\varphi} \text{Re}\left( 
       \int^{\,\text{e}^{u+\text{i}\varphi}}_0\!\!\! Q(t)\,\text{d}t\right),
     \label{WKB_ControllingFactor}
\end{equation}
where $Q(t)$ is found from the differential equation~(\ref{ODE}).
We then have the relation
\begin{align}
  S(u) &= \log\left(\left| a_{\bar{m}}\right|\right) + \left(\nu+\bar{m}\right)u,\label{SuDefinition}\\
  u &= \left.-\frac{d}{d m} \log\left(\left| a_{m}\right|\right)\right|_{m=\bar{m}}.\label{uDefinition}
\end{align}
The last equation follows from the maximum condition. We recognize
(\ref{SuDefinition}, \ref{uDefinition}) as a Legendre
transform~[\cite{LegendreTransform1},\cite{LegendreTransform2}, \cite{LegendreTransform3}].
By inverting this transform we find
\begin{align}
   \bar{m} = \frac{d}{d u} S(u),\quad
   \log\left(\left| a_{\bar{m}}\right|\right) = S(u) - \left(\nu+\bar{m}\right)u,
   \label{LegendreTransform}
\end{align}
which provides an {\em a priori\/}
order-of-magnitude estimate of the coefficients $a_m$,
and hence of (i) the accuracy loss due to numerical roundoff,
and (ii) the number of terms ${\cal M}$ required in (\ref{SeriesExpansion})
for a desired final precision.

\subsection{Example 1: Anharmonic oscillators}

Consider the equation
\begin{equation}
   -\frac{\partial^2}{\partial y^2}\Psi(y) + \left(y^2+ c^2\right)^2\Psi(y) = 0.
   \label{Anharmonic_oscillator}
\end{equation}
For large $y$ the typical solution behaves like
\begin{equation}
     \Psi(y) \sim \text{e}^{\frac{1}{3}y^3 + c^2 y},
\end{equation}
neglecting the slowly varying prefactor. Equation (\ref{Anharmonic_oscillator})
can be transformed to the form (\ref{ODE}) by introducing $x=y^2$, 
$\Psi(y) = \psi(x)$. Hence, with $x=y^2=\text{e}^{u}$
\begin{equation*}
     S(u) = {\textstyle \frac{1}{3}}\left(\text{e}^{\frac{3}{2}u} + 3 c^2 \text{e}^{\frac{1}{2}u}\right),
\end{equation*}
which gives
\begin{align}
  \bar{m} &= {\textstyle \frac{1}{2}}\left(\text{e}^{\frac{3}{2}u} + c^2\,\text{e}^{\frac{1}{2}u} \right),
  \label{m_Anharmonic}
  \\
  \log\left(\left| a_{\bar{m}}\right|\right) &= 
  \left({\textstyle \frac{1}{3}} - {\textstyle \frac{1}{2}} u\right) \text{e}^{\frac{3}{2}u} + 
  c^2 \left(1 -{\textstyle \frac{1}{2}} u \right) \text{e}^{\frac{1}{2}u}.
  \label{a_m_Anharmonic}
\end{align}

\begin{figure}[H]
\begin{center}
\includegraphics[clip, trim = 8ex 6ex 9ex 5ex, width=0.483\textwidth]{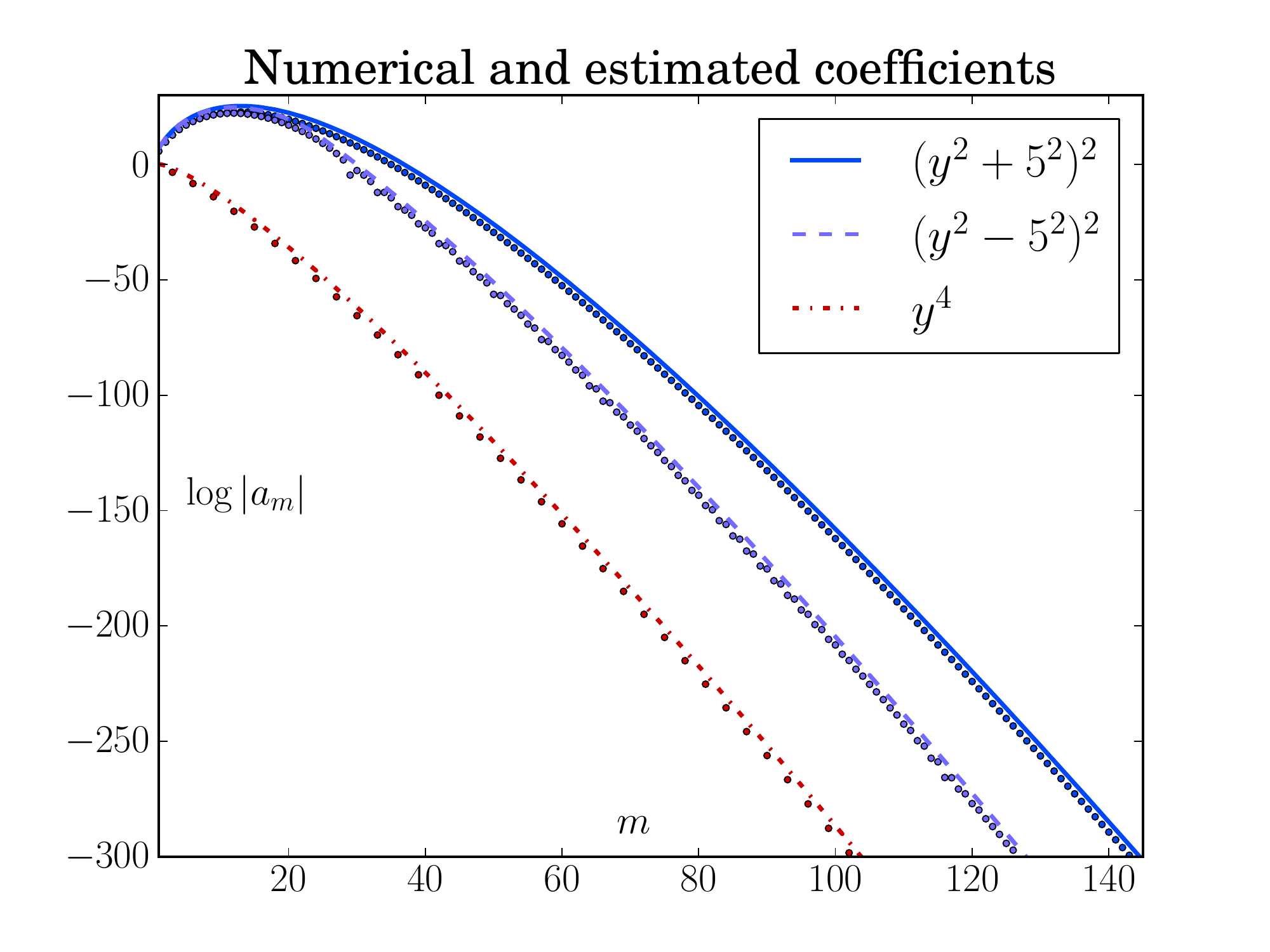}
\end{center}
\caption{Comparison of numerical coefficients $a_m$ (points)
with estimates (full-drawn lines) based on
(\ref{m_Anharmonic}, \ref{a_m_Anharmonic}) and
(\ref{m_DoubleWell}, \ref{a_m_DoubleWell}). The estimates
of $\log\vert a_m \vert$ are accurate up to corrections which
depend logarithmically on $m$.
}
\label{Coefficients_a_m}
\end{figure}

Note that (\ref{m_Anharmonic}, \ref{a_m_Anharmonic}) give, in parametric form,
an estimate of {\em all\/} coefficients $\vert a_m \vert$,
not only those corresponding to a maximum value. For $c=0$ an explicit representation
is easily found to be
\begin{equation}
   \log \vert a_m \vert = \frac{2}{3}m\left(1 -\log 2m \right).
   \label{cZeroCoefficients}
\end{equation}
This is plotted as the lower curve in figure~\ref{Coefficients_a_m}. It fits satisfactory
with the high-precision coefficients generated numerically, but there remains a
correction which depends logarithmically on $m$. For nonzero $c$ the parametric
representation provides equally good results, as shown by the upper curve in 
figure~\ref{Coefficients_a_m}.

\begin{figure}[H]
\begin{center}
\includegraphics[clip, trim = 10.5ex 5ex 10ex 5ex, width=0.483\textwidth]{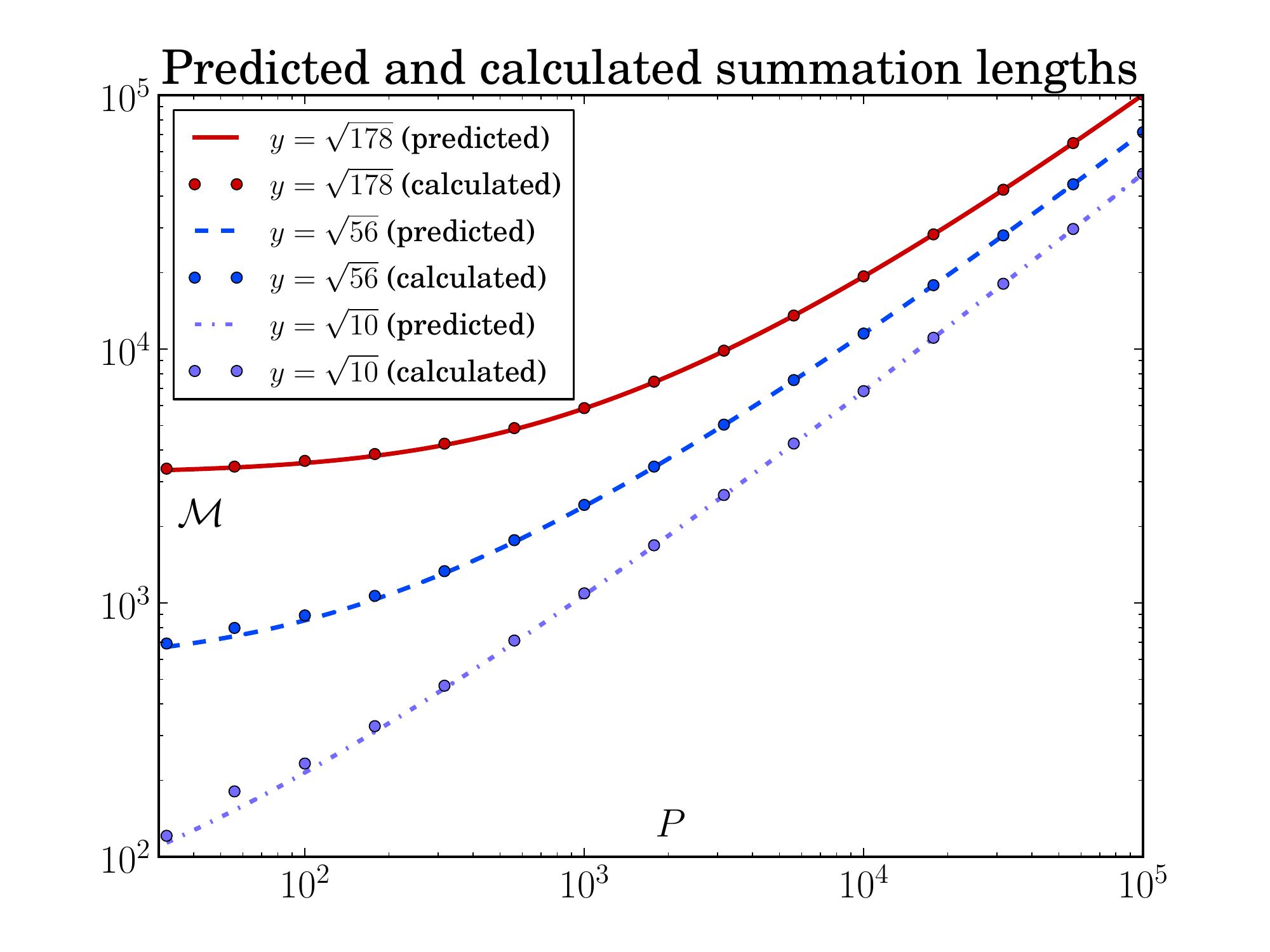}
\end{center}
\caption{This figure compares the {\em a priori\/} prediction,
based on equation~(\ref{cZeroCoefficients}),
of the number of terms ${\cal M}$ which must be summed in order to evaluate
$\Psi(y)$ for $c=0$ to a desired precision $P$ with the actual number of terms
computed by \lstinline!seriesSolveOde1!.
}
\label{lengthOfSums}
\end{figure}

The conclusion of this example is that we expect the largest term of the power series to be
\begin{equation}
    \mathop{\text{max}}_m \vert A_m(x) \vert \sim \text{e}^{\frac{1}{3}(x^{3/2}+3 c^2 x^{1/2})},
\end{equation}
neglecting a slowly varying prefactor.
Further, the maximum should occur at
\begin{equation}
    m \approx {\textstyle \frac{1}{2}} \left( x^{3/2} + c^2 x^{1/2} \right).
\end{equation}
Finally, estimates like equation (\ref{cZeroCoefficients})
for the coefficients $a_m$ may be used to predict how many terms ${\cal M}$
we must sum to evaluate $\psi(x)$ to a given precision $P$,
based on the stopping criterium
\begin{equation}
       \vert a_{\cal M} \vert\, x^{\cal{M}} \le 10^{-P}.
       \label{stoppingCriterium}
\end{equation}
As can be seen in figure~\ref{lengthOfSums} the agreement with the actual
number of terms used by \lstinline!seriesSolveOde1! is good,
for case of equation~(\ref{Anharmonic_oscillator}) with $c=0$,
in particular for high precision $P$. But note that a
logarithmic scale makes it easier for a comparison to look good.

\subsection{Example 2: Double well oscillators}

Next consider the equation
\begin{equation}
   -\frac{\partial^2}{\partial y^2}\Psi(y) + \left(y^2- c^2\right)^2\Psi(y) = 0.
   \label{Double_well}
\end{equation}

For large $y$ the typical solution behaves like
\begin{equation}
     \Psi(y) \sim \text{e}^{\frac{1}{3}y^3 - c^2 y},
\end{equation}
neglecting the slowly varying prefactor. Equation (\ref{Double_well})
can be transformed to the form (\ref{ODE}) by introducing $x=y^2$, 
$\Psi(y) = \psi(x)$. Hence, with $x=y^2=\text{e}^{u}$
\begin{equation*}
     S(u) = \mathop{\text{max}}_\varphi  {\textstyle \frac{1}{3}}\text{Re} \left(\text{e}^{\frac{3}{2}(u+\text{i}\varphi)} 
       - 3 c^2 \text{e}^{\frac{1}{2}(u+\text{i})\varphi}\right).
\end{equation*}
The maximum occurs for $\cos\frac{1}{2}\varphi = -\frac{1}{2}\left(1 + c^2\,\text{e}^{-u} \right)^{1/2}$ when
$\text{e}^{u} \ge \frac{1}{3} c^2$, and for $\cos\frac{1}{2}\varphi=-1$ otherwise.
This gives
\begin{equation}
     S(u) = \left\{\begin{array}{cc}
       {\textstyle c^2 \text{e}^{u/2} - \frac{1}{3}\text{e}^{3u/2}}&\text{for $e^u \le \frac{1}{3}c^2$,}\\[0.5ex]
       {\textstyle \frac{1}{3}} (\text{e}^{u} + c^2)^{3/2}&\text{for $e^u \ge \frac{1}{3}c^2$.}
       \end{array}
       \right.
\end{equation}
This implies that
\begin{align}
     \bar{m} &= 
     \left\{\begin{array}{lc}
         \frac{1}{2} \text{e}^{u/2}\left(c^2 - e^u\right)
         &\text{for $e^u \le \frac{1}{3}c^2$,}\\[0.5ex]
     {\textstyle \frac{1}{2}} \text{e}^{u}\,\left( \text{e}^u + c^2 \right)^{1/2}&\text{for $e^u \ge \frac{1}{3}c^2$},
     \end{array}
     \right.
     \label{m_DoubleWell}
     \\
     \log\left(\left| a_{\bar{m}}\right|\right) &= 
     \left\{\begin{array}{cc}
         \left(1-\frac{1}{2}u\right)c^2 \text{e}^{u/2} -\left(\frac{1}{3}-\frac{1}{2}u\right)\text{e}^{3u/2}
        &\text{for $e^u \le \frac{1}{3}c^2$,}\\[0.5ex]
     \left[\left({\textstyle \frac{1}{3}} -{\textstyle \frac{1}{2}}u \right)\text{e}^{u} 
       +{\textstyle \frac{1}{3}}c^2\right]\left(\text{e}^u + c^2\right)^{1/2}&\text{for $e^u \ge \frac{1}{3}c^2$}.
     \end{array}
     \right.
     \label{a_m_DoubleWell}
\end{align}
This representation compares fairly well with the numerically generated coefficients,
as shown by the middle curve in figure~\ref{Coefficients_a_m}. However, in this case
the coefficients $a_m$ have a local oscillating behaviour. The representation
(\ref{m_DoubleWell}, \ref{a_m_DoubleWell}) should be interpreted as the local amplitude
of this oscillation.

The conclusion of this example is that we expect the largest term of the power series to be
term of the series to be
\begin{equation}
    \mathop{\text{max}}_m \vert A_m(x) \vert \sim \text{e}^{\frac{1}{3}(x + c^2)^{3/2}},
\end{equation}
neglecting the slowly varying prefactor.
Further, the maximum should occur at
\begin{equation}
    m \approx {\textstyle \frac{1}{2}} x \left( x + c^2 \right)^{1/2} 
    \approx {\textstyle \frac{1}{2}} x^{3/2} + {\textstyle \frac{1}{4}} c^2 x^{1/2}.
\end{equation}

\subsection{Logarithmic corrections}

\begin{figure}[H]
\begin{center}
\includegraphics[clip, trim = 8ex 5ex 9ex 5ex, width=0.483\textwidth]{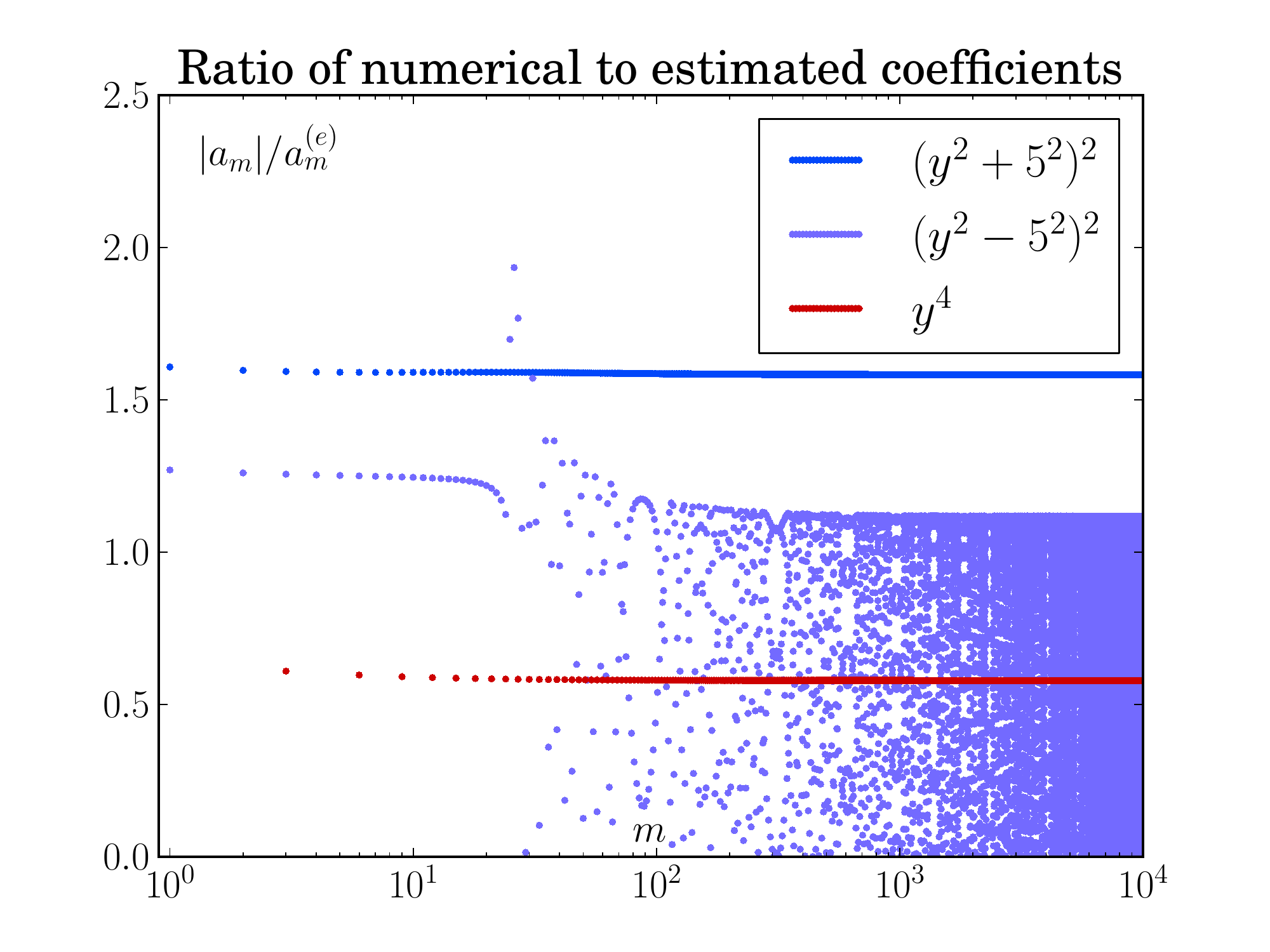}
\end{center}
\caption{The ratio between the numerically generated $\vert a_m \vert$ and their
estimated values $a^{(e)}_m$ based on equation~(\ref{LegendreTransform}) with $S(u)$ replaced
by $S_\text{eff}(u)$. The important feature here is that the ratios are essentially constant,
not which constant, due to an overall normalization constant which we have not attempted
to estimate here.
}
\label{coefficientRatio}
\end{figure}

The agreement between the {\em a priori\/} estimated
magnitude $a^{(e)}$ and the numerically generated
values $\vert a_m \vert$ looks quite good in figure~\ref{Coefficients_a_m}.
This is partly due to the logarithmic scale; on closer examination
the coefficients are seen to differ by many orders of magnitude. The
agreement can be significantly improved by (i) taking into account the
prefactor $Q(u)^{-1/2}$ in the WKB-expression (\ref{WKB_ControllingFactor})
for the left hand side of (\ref{BasicHypothesis}), and 
(ii)  changing the right hand side of (\ref{BasicHypothesis}) as
\begin{align*}
    \vert a_{\bar{m}}\, \vert x^{\nu +\bar{m}} \equiv \text{e}^{s(\bar{m}) + (\nu + \bar{m})u}
    \to \sqrt{{2\pi\, S''(u)}}\,\text{e}^{s(\bar{m}) + (\nu + \bar{m})u}.
\end{align*}
The latter replacement takes into account that the main contribution to the
sum over $m$ comes from a range of values around $\bar{m}$,
approximates this contribution by a gaussian integral,
and uses the fact that $s''(\bar{m})^{-1} = -S''(u)$.
These two improvements amounts to a change $S(u) \to S_{\text{eff}}(u)$.

Implementing these changes for the $c=0$ case
of (\ref{Anharmonic_oscillator}), taking into account that
only coefficients $a_{3n}\ne 0$,
the estimate (\ref{cZeroCoefficients}) can be improved to
\begin{equation}
{\textstyle
   \log a^{(e)}_m = \frac{2}{3}\left(m + {\textstyle \frac{5}{4}}\right)
   \left[1 -\log(2m + \frac{5}{2})\right]
   - \frac{1}{2}\log\frac{\pi}{6}.
}
   \label{cZeroImproved}
\end{equation}
Similar, algebraically more complicated, improvements can be made for $c>0$.
As shown in figure~\ref{coefficientRatio} the improved estimates compare
very well with the numerically generated coefficients, but in case of locally
oscillating $a_m$ the (smooth) estimate $a^{(e)}_m$ should be interpreted as
the local oscillation amplitude (as illustrated by the $(y^2-5^2)^2$--case
in figure~\ref{coefficientRatio}). 

\section{Concluding remarks}

For equations without singular points in the finite plane our code
can be used for expansion around any point $\zeta_0$ in the
complex plane. To shift from one expansion point to another one
just has to rewrite the parameters $\text{v}_n$,
and let $z$ denote the distance from $\zeta_0$. This allows for analytic
continuation of the solution, which becomes quite easy since the full solution
is determined by just the two parameters $\psi(\zeta_0)$, $\psi'(\zeta_0)$
(in addition to the differential equation).

The strategy of using a sequence of series expansions,
each with a small parameter $z$, has been used by
Haftel {\em et al}  \cite{HaftelKrivecMandelzweig}. The
advantage is that each summation requires fewer terms in
the series, and may lead to less loss of precision caused
by roundoff errors. The cost is of course that one has to
do several sums, and one may also loose symmetries like the
$y\to -y$ symmetry in equations  (\ref{AnharmonicOscillator}, 
\ref{DoubleWell}). The latter leads to more algebraic
operations per recursion step.

The optimal strategy may depend on the problem. If we are
only solving equation (\ref{AnharmonicOscillator}) for the
ground state eigenvalue $\varepsilon_0$, this is basically
determined by the condition that the asymptotic behaviour of
the solution switches very rapidly between $\text{e}^{y^3/3}$ and
$-\text{e}^{y^3/3}$. This behaviour is not affected much by roundoff errors.
Consider the question is whether it is faster to evaluate $\text{e}^{y^3/3}$ by a
single series expansion, or by $k$ expansions
with $y_k = y/k$. By combining equations (\ref{cZeroCoefficients}, \ref{stoppingCriterium})
one finds that each sum requires about ${\cal M}_k$ terms
for a given precision $P$, where ${\cal M}_k$ satisfies
the equation
\begin{equation}
   \frac{2}{3}{\cal M}_k\left(1-\log2{\cal M}_k\right) 
   + 2 {\cal M}_k\log({y}/{k}) \approx - P \log 10,
\end{equation}
which is best solved numerically. Consider f.i.~the case of
$y=\sqrt{178}$ and $P = 10^{5}$. As can be seen from
figure~\ref{lengthOfSums} about ${\cal M} \equiv {\cal M}_1 = 10^5$ terms
have to be summed to obtain the desired precision. With $k=2$ only about
${\cal M}_2 = 67\,500$ terms has to be summed, but since this has to be done
twice the total effort becomes larger. The situation is similar for other
values of $y$ and $P$.

In other cases, like highly excited states of (\ref{AnharmonicOscillator}) 
or all states of (\ref{DoubleWell}), there is a loss of precision
due to roundoff. This changes how the number of terms ${\cal M}$
and the actual precision $M$ vary with $y$. The latter is most
important since the time per multiplication increases somewhat
faster than quadratic with $M$. In such cases a sequence of analytically
continued evaluations are clearly advantageous; optimization of the
number and size of steps requires some prior knowledge of the coefficients
$a_n$ and the behaviour of the solution \cite{ANKO_CPC2012}.
If one needs to evaluate the solution at a sequence of points, as
when calculating the normalization integral \cite{ANKO}, analytic
continuation would also be preferrable.

The routine \lstinline!seriesSolveOde1! does not allow for analytic
continuation in the presence of a regular singular point, since the
transformed equation belong to a different class. We have developed
and are testing code for a more general class of equations
(as hinted by our naming scheme), which we intend to submit real soon.
This code allow for translations (or more generally M{\"o}bius
transformations) to a new expansion point $\zeta_0$. It can f.i.~be
used to solve Mathieu and Mathieu-like equations.

We have made extensive tests of the submitted code, which appears to be 
robust and perform according to theoretical expectations. As illustrated, 
surprisingly accurate {\em a priori\/} estimates of the series to be summed 
can be made by using the WKB approximation in combination with Legendre transforms. 
This is useful for estimating precision and time requirements in advance. 
In the general case the WKB integral and Legendre transformation must be computed
numerically. We have developed and tested code for this purpose \cite{ANKO_London2012, ANKO_CPC2012}.

\section*{Acknowledgement}
\label{acknowledgement}
We thank A.~Mushtaq and I.~{\O}verb{\o} for useful discussions.
This work was supported in part by the Higher Education Commission of Pakistan (HEC). 


\newpage

\section*{Appendix}

\newcommand\abs{\text{abs}}
\renewcommand\Re{\text{Re}}
\renewcommand\Im{\text{Im}}

{\footnotesize
\lstset{
  language=C++,
  title={\normalsize Definition of the OdeParams structure},
  morekeywords={*, cl_N, cl_I, float_format_t}
}
\begin{lstlisting}
struct OdeParams {
  // Parameters defining the ODE
  cl_N  z;             // Evaluation point
  cl_N  nuP;           // $\nu_+$
  cl_N  nuM;           // $\nu_-$
  cl_N  s;
  int   orderN;        // Order of polynomial potential
  cl_N* v;             // Pointer to array v[orderN+1]
  // Options regulating the computation
  float_format_t prec; // Wanted computational precision
  bool  nuPlus;        // If true, compute $\nu=\nu_+$ solution
  bool  dAlso;         // If true, compute derivative also
  cl_I  emmMax;        // Stop summation when emm=emmMax
  cl_I  emmTooLarge;   // Stop if emm >= emmTooLarge
  // Options to print double precision coeff info to stdout
  bool  writeParams;   // Write to (temporary) file?
  bool  deleteParams;  // Delete above file at normal exit?
  bool  printLogAbsA;  // Option to print all $\log(\abs(A_m))$
  bool  printLogReA;   // Option to print $\log(\abs(\Re(A_m)))$
  bool  printLogImA;   // Option to print $\log(\abs(\Im(A_m)))$
  bool  printArgA;     // Option to print $\arg(A_m)$
  bool  printReA;      // Option to print $\Re(A_m)$
  bool  printImA;      // Option to print $\Im(A_m)$
  bool  printLogAbsAd; // Print all $\log(\abs((\nu+m)A_m/z))$ 
  bool  printLogReAd;  // Print $\log(\abs(\Re((\nu+m)A_m)/z))$
  bool  printLogImAd;  // Print $\log(\abs(\Im((\nu+m)A_m/z)))$
  bool  printArgAd;    // To print $\arg((\nu+m)A_m)/z$
  bool  printReAd;     // To print $\Re((\nu+m)A_m/z)$
  bool  printImAd;     // To print $\Im((\nu+m)A_m/z)$
  // Options to print full precision coefficients to stdout 
  bool  coutA;         // Option to print $A_m$
  bool  coutReA;       // Option to print $\Re(A_m)$
  bool  coutImA;       // Option to print $\Im(A_m)$
  bool  coutAd;        // Option to print $(\nu+m)A_m/z$
  bool  coutReAd;      // Option to print $\Re((\nu+m)A_m/z)$
  bool  coutImAd;      // Option to print $\Im((\nu+m)A_m/z)$
};
\end{lstlisting}
}

{\footnotesize
\lstset{
  language=C++,
  title={\normalsize Definition of the OdeResults structure},
  morekeywords={*, cl_N, cl_I, float_format_t}
}
\begin{lstlisting}
struct OdeResults {
  cl_N* fu;          // Pointer to pre-created array fu[2]
  int maxAExponent;  // Largest term in function series
  int maxAdExponent; // Largest term in derivative series
  int emmAtAMax;     // $m$ where $\abs(A_m)$ is largest
  int emmAtAdMax;    // $m$ where $\abs((\nu+m)A_m)$ is largest
  int lengthOfSum;   // Number of terms summed
  double timeUsed;   // Time used for evaluation
  double lgErrorF;   // Estimated error in function
  double lgErrorFd;  // Estimated error in derivative
  int returnStatus;  // -1: $s=0$
                     // -2: $z=0$
                     // -3: $(\nu_P - \nu_M)$ is integer $!= 1$
                     // -4: emm >= emmTooLarge was reached
                     //  1: All $\abs(A_m) < \text{Estimated error}$
                     //  2: emm = emmMax was reached
};
\end{lstlisting}
}

\end{document}